# Grain Dependent Growth of Bright Quantum Emitters in Hexagonal Boron Nitride

Noah Mendelson,[1,*] Luis Morales,[2] Chi Li,[1] Ritika Ritika,[1] Minh Anh Phan Nguyen,[1] Jacqueline Loyola-Echeverria,[1] Sejeong Kim,[1] Stephan Götzinger,[2,3,4] Milos Toth,[1,5] and Igor Aharonovich[1,5,*]

[1]School of Mathematical and Physical Sciences, University of Technology Sydney, Ultimo, New South Wales 2007, Australia.
[2]Max Planck Institute for the Science of Light, D-91058 Erlangen, Germany.
[3]Department of Physics, Friedrich Alexander University Erlangen-Nuremberg (FAU), D-91058 Erlangen, Germany.
[4]Graduate School in Advanced Optical Technologies (SAOT), FAU, D-91058 Erlangen, Germany.
[5]ARC Centre of Excellence for Transformative Meta-Optical Systems, University of Technology Sydney, Ultimo, New South Wales, Australia.

*Noah.B.Mendelson@student.uts.edu.au
*Igor.Aharonovich@uts.edu.au

**Abstract:**
Point defects in hexagonal boron nitride have emerged as a promising quantum light source due to their bright and photostable room temperature emission. In this work, we study the incorporation of quantum emitters during chemical vapor deposition growth on a nickel substrate. Combining a range of characterization techniques, we demonstrate that the incorporation of quantum emitters is limited to (001) oriented nickel grains. Such emitters display improved emission properties in terms of brightness and stability. We further utilize these emitters and integrate them with a compact optical antenna enhancing light collection from the sources. The hybrid device yields average saturation count rates of ~2.9 x$10^6$ cps and an average photon purity of ~90%. Our results advance the controlled generation of spatially distributed quantum emitters in hBN and demonstrate a key step towards on-chip devices with maximum collection efficiency.

**Introduction:**
Atomic defects in solid state hosts that act as single photon emitters (SPEs) are promising hardware components for a range of emerging quantum technologies,[1] such as quantum network nodes, quantum repeaters, and quantum key distribution. Despite rapid advances culminating in these proof of principle demonstrations, reliable fabrication techniques and subsequent integration with nanophotonic components remains challenging.[2]

Recently, layered van der Waals materials such as hexagonal boron nitride (hBN) have been explored as hosts of SPEs.[3, 4] This class of materials is of a particular interest as their layered nature provides avenues for nanoscale manipulation and reliable integration with photonic devices.[5-8] hBN SPEs are especially promising due to the stability of the host,[9] room temperature operation,[10] outstanding optical properties,[11] and the presence of optically detected magnetic resonance signals.[12-14]

SPEs in hBN can be engineered through chemical vapor deposition (CVD),[15-17] molecular beam epitaxy,[12, 18] or metal organic vapor phase epitaxy.[12] Bottom up fabrication encompasses two primary advantages over top down fabrication methods. The first is the ability to fabricate cm scale thin films with controlled thickness that are ideal for transfer and incorporation with nanophotonic platforms.[19] The second is the ability to manipulate the photophysical properties of the incorporated SPEs during growth, such as defect density,[20] emission energy,[20] and the fabrication of homogeneous ensembles.[12]

In this work, we report on low-pressure CVD growth of hBN on polycrystalline Ni foils. We demonstrate that SPEs are selectively incorporated only in hBN grown on (001) Ni, despite confirming that growth occurs on all grain orientations. Remarkably, the SPEs incorporated on (001) Ni display superior emission intensity and stability compared to previous CVD growth studies on copper.[15-17] Consequently, we transfer the few-layer hBN to a planar dielectric antenna to enhance its collection efficiency. Our hybrid device yields average saturated count rates of $\sim 3 \times 10^6$ counts/s and a high emission purity $g^2(0) \sim 0.1$. Our results advance the understanding of SPE formation in layered materials and provide a unique and promising route to deterministic SPEs for quantum photonics.

Figure 1a displays a schematic illustration of the CVD hBN growth on polycrystalline Ni foil. The growth of hBN *via* low-pressure CVD was performed following an established protocol.[15] Figure 1b displays a Fourier-transform infrared spectrum recorded from as-grown multilayer hBN films. The spectrum is from a large area that contains all Ni grain orientations, and displays the three characteristic FTIR active stretching modes,[21] confirming the growth of few-layer hBN. Figure 1c displays an optical image of a selected region of the Ni film covered with few-layer hBN. A clear contrast in the colors of the different Ni grains can be observed, consistent with previous reports.[22] Figure 1d displays the same region analyzed by scanning electron microscopy, where the grain boundaries are similarly visible.

To determine the orientation of each Ni grain, we analyzed the region with electron beam secondary diffraction (EBSD), Figure 1e. The image displays the orientation of the Ni domains at the surface, and it can be seen that (111), (101), and (001) oriented grains are distributed randomly across the foil with individual grains showing surfaces areas of $\sim 10\text{-}50\mu m^2$. To probe for local variations in the material quality of the as-grown hBN film (i.e. within a particular single crystal grain), we collected the Raman spectra from (111), (101), and (001) oriented grains displayed in Figure 1f. hBN displays only one prominent Raman active stretching mode – an $E_{2g}$ mode at $\sim 1367 cm^{-1}$.[23] Note that the collection parameters were held constant across the measurements, and the individual peaks were normalized to the (001) spectrum, permitting a direct comparison of relative intensity. For (101) and (001) grains we observe a strong resonance at $1368.1 cm^{-1}$ and $1367.5 cm^{-1}$, respectively. The signal is $\sim 20\%$ weaker on the (101) grain, while both display a near-identical peak full width at half maximum (FWHM) of $\sim 33.5 cm^{-1}$. The similar widths suggest similar material quality,[24] while the hBN film may be slightly thicker on the (001) grains.[23] In contrast, we only observe a weak Raman signal for hBN grown on (111) oriented Ni grains. The presence of the $E_{2g}$ mode centered at $1367.4 cm^{-1}$ confirms that hBN is present on (111) grains. The signal is $\sim 5$ times weaker, suggesting that less material has been grown. However, the peak displays a narrower FWHM ($18.7 cm^{-1}$) than observed on the (101) and (001) grains, indicating a slightly higher material quality of the hBN grown on (111) grains. We further discuss the implications of the grain dependent growth rate, and underlying mechanism below.

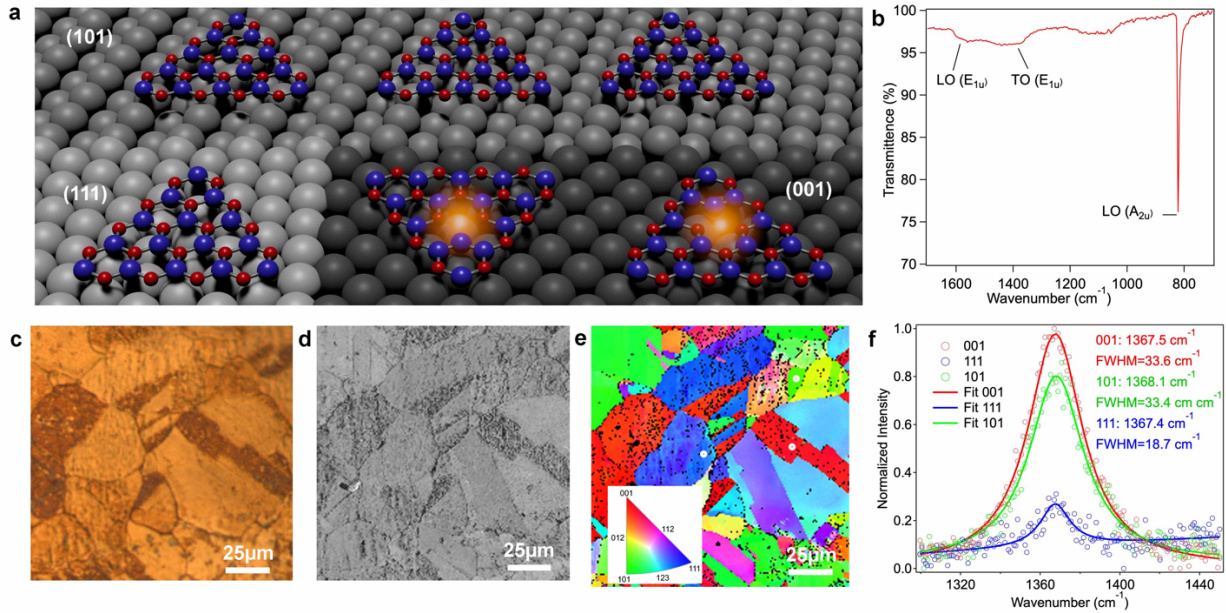

*Figure 1. CVD growth of hBN on polycrystalline Ni foil. **a.** Schematic illustration of hBN growth on Ni (111), (101), and (001) grains, where orange spheres represent SPEs which are only observed on (001) oriented Ni grains. **b.** FTIR spectrum from the as-grown hBN on nickel, showing the LO and TO active modes. Optical **(c)** SEM **(d)** and EBSD **(e)** images of as-grown hBN on Ni. Inset in e displays the legend relating the grain orientation to color. **f.** Raman spectroscopy of hBN on different Ni grains (111), (101), (001). The spot of each collection is marked with a white circle in panel e.*

We next characterize the photoluminescence (PL) of hBN grown on different Ni grains to examine potential variations in SPE incorporation. Note that all measurements were performed with the hBN adhered to the Ni growth substrate. Figure 2a displays a selected region where the grains are sufficiently small to be investigated in-tandem, within the field of view of our optical setup. Utilizing a custom-built scanning confocal PL setup, we mapped a selected area (indicated by the black dashed line in figure 2a). Figure 2b displays the resulting confocal scan, where we observe a clear dependence of emission intensity from the hBN film on the orientation of the underlying Ni substrate. Specifically, hBN on (001) grains displays a bright fluorescence, whilst that on (101) and (111) grains is drastically reduced.

We further checked for the presence of SPEs in each subsequent grain orientation, by recording the spectrum from localized excitation spots. Figure 2c shows a representative spectrum of an SPE located in a (001) Ni grain, displaying the characteristic line shape of hBN SPEs in the visible spectral range. The emitter has a zero-phonon-line (ZPL) at ~573nm and a phonon side-band (PSB) consisting of 2 peaks at ~644nm and ~653nm.[25] To probe the nature of the emission we performed second order auto-correlation measurements on the emission peak, and obtained $g^2(0) < 0.5$, confirming the quantum nature of the emission, inset Figure 2c. While SPEs are easily located within corresponding (001) Ni grains, we were unable to locate any narrow emission peaks resembling hBN SPEs in either (101) or (111) Ni grains within the entire 125x125μm scan area in Figure 2b.

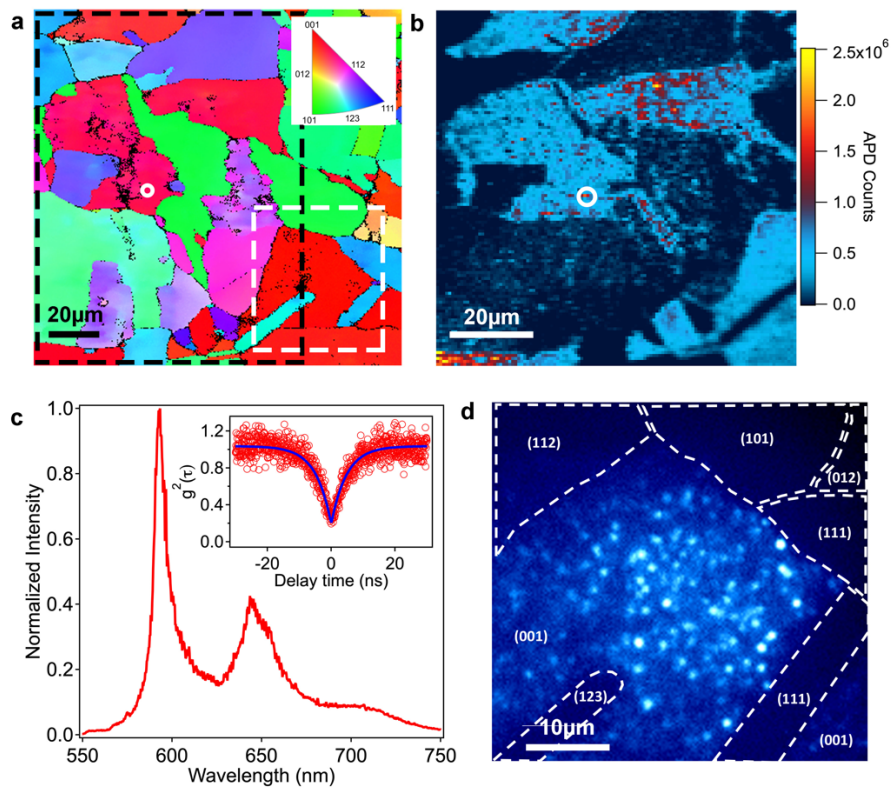

*Figure 2. Grain dependent incorporation of hBN SPEs. **a.** EBSD map demonstrating the Ni grain orientation in the characterized area. The black (white) dashed box corresponds to the area mapped in b (d) respectively, white the white circle marks the SPE in c. **b.** Confocal PL mapping (532nm 300µW) of the black dashed area. The white circle marks the SPE displayed in c. **c.** Spectrum of representative SPE localized at the white circle in a and b. Inset displays the $g^2(\tau)$ from the same emitter. **d.** Wide-Field imaging of white dashed area marked in a. A high density of SPEs on (001) Ni grains is apparent while no SPEs appear in alternative grain orientations.*

To directly visualize the grain dependent incorporation of SPEs and to evaluate the density of SPEs in (001) grains we additionally performed wide-field imaging, as shown in figure 2d. The imaged area corresponds to the white dashed area in figure 2a. Each localized emission center is attributed to an SPE. These results confirm the high density of SPEs incorporated in (001) Ni grains, while SPE are not observed on alternative grain orientations. Additional areas characterized *via* wide-field imaging all demonstrate that incorporation of SPEs is limited to (001) oriented grains, see Supporting Information.

In an attempt to understand the selective incorporation of SPEs on Ni (001) grains, we reflect on current understanding of the growth mechanism of hBN on Ni. It has previously been postulated that variations in the surface energy,[22] or the sticking coefficient of the precursor compounds,[26] on different Ni grain orientations is responsible for the observed growth rate dependencies. However, more recent work has confirmed that the role of diffusion and segregation of precursor species within the Ni catalyst also play a central role,[27] which is supported by the understanding that diffusion kinetics of atomic species such as B, C, and N within Ni are highly sensitive to temperature as well as the grain orientation of the Ni.[28, 29] In fact, at temperatures near that used during growth in this study (1030°C), the diffusivity of atoms within nickel is highest on the (001) plane.[28, 29] Interestingly, previous studies have demonstrated that modifying catalyst diffusion effects can control both the density and emission energies of hBN SPEs during CVD growth on copper.[20]

As a result, we propose that grain-dependent variations in atomic diffusion processes provide the most straightforward interpretation of the selective incorporation of SPEs on Ni (001) grains. The consequences of the proposed mechanism are twofold. The first is through modified feeding rates of B and N to the surface, a parameter known to influence the formation energies of atomic defects.[30] The second is *via* the incorporation dynamics of heteroatom impurities such as carbon, which has recently been linked to SPEs in hBN.[12] Our interpretation is strengthened by noting that while grain dependent diffusion effects are also observed in CVD of hBN on Cu, they are relatively minor in comparison to Ni showing negligible variations in the observed growth kinetics.[31] This is consistent with previous reports demonstrating that SPEs are incorporated homogeneously across various Cu grain types,[15] suggesting grain based diffusion variations are not significant enough to affect SPE incorporation on Cu.

Having established the grain dependent incorporation of SPEs on Ni (001), we explore in further depth their optical properties. Figure 3a displays a histogram of ZPL energies for 84 separate SPEs localized on Ni (001). Emitter ZPLs are found to be clustered around 580 nm, with ~87% <600nm, in agreement with previous studies on SPEs in CVD grown hBN materials.[15, 16] The inset for Figure 3a shows a typical stability trace for a Ni (001) emitter at an excitation power of 300μW, demonstrating a stable emission rate despite the nanoscale dimensions of the film (~10 nm)

Next, we compare the overall intensity of the emitters grown on Ni and Cu foils. Figure 3b (bottom panel) plots the emission intensities of 10 representative SPEs characterized on Ni (001) grains (blue) contrasted with 10 SPEs grown on Cu foil (red). Each sample was grown under otherwise identical growth conditions, and their brightness compared by evaluating the relative spectrometer counts under equivalent excitation conditions (300μW, 10sec acquisition). Brightness was evaluated *via* spectral comparison (as opposed to photon counting) to reduce the potential contribution of varying background emission rates between the samples. SPEs incorporated on Ni (001) display an average ~2.5-fold enhancement in emission compared to those on copper (dashed lines Figure 3b). The enhancement fluorescence of SPEs on Ni (001) planes is likely the result of favorable environmental influences—such as a decreased in the prevalence of nearby charge traps or additional point defects—as quantum efficiency and brightness are known to vary even locally,[32, 33] however, the exact mechanism responsible for emission rate fluctuations in hBN SPEs remains unknown. Figure 3b (top panel) displays the three brightest emitters characterized on Ni (001) grains, the brightest of which displayed a fluorescence signal roughly 10x the average SPE observed on Ni (001) with an estimated count rate of ~$7.6*10^6$ cps, see Supporting Information. The extreme brightness of this particular SPE highlights the potential for ultra-bright emission centers fabricated *via* bottom up growth.

The improved emission rate and stability of SPEs incorporated on Ni (001) grains encompassed within a nanometer scale host material (<10nm) makes this system appealing for integration into planar dielectric antennas (PDAs) which can enhance the collection efficiency from the source.[34] PDA structures have been successfully implemented in the past for efficient light extraction from single molecules in organic crystals,[35] quantum dots in thin polymer films,[36] and Nitrogen vacancy centers in diamond.[37] These devices are inherently broadband,[38] robust against fabrication imperfections, and compatible with a variety of single photon sources.

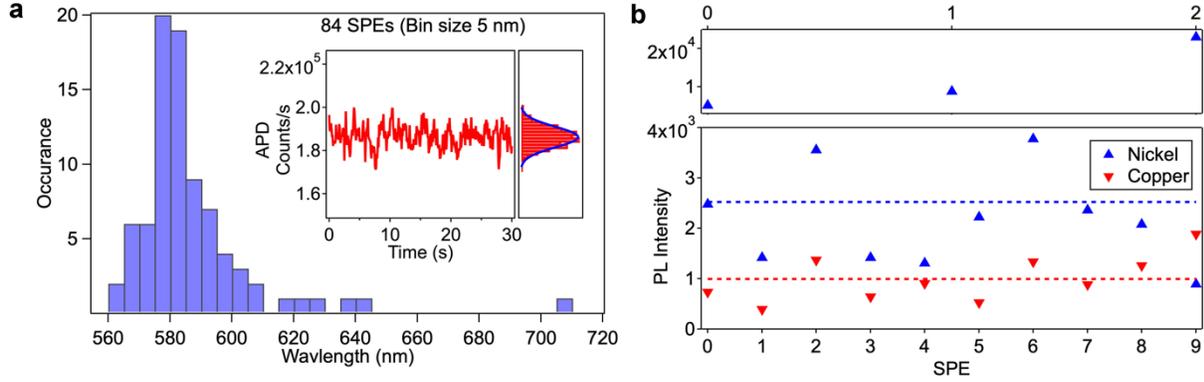

*Figure 3. Optical properties of hBN SPEs incorporated on Ni (001) grains. a. Histogram plotting the ZPL positions of 84 analyzed SPEs on Ni (001) grains, with a 5nm bin width. The emission lines are clustered around 580nm. The inset shows a representative stability trace from an SPE at 300µW b. The lower panel is a scatter plot comparing the relative brightness of 10 representative SPEs on Ni (001) grains (blue triangles) and on Cu (red triangles). Dashed lines indicate the average count rates for each set of 10 SPEs. The upper panel records the brightest SPEs observed from Ni (001) grains, which are separated due to their higher relative brightness.*

Figure 4a depicts a schematic illustration of the PDA design used in this work, consisting of a glass coverslip capped with a 250 nm layer of $MgF_2$ on top of which the hBN is transferred (cf. methods). To ensure optimal operation of the hybrid structure, we must consider the refractive index (n), the thickness of the hBN layer, as well as the emission energy ($\lambda_{SPE}$) of the SPE. Specifically, for a hBN layer thickness ($n_{hBN}$=1.8)[39] approaching $\lambda_{SPE} / (2 * n_{hBN})$, the appearance of propagating modes within hBN will occur. In this case the hBN layer will act as a 2D waveguide created by the $MgF_2$/hBN/air stack ($n_{MgF2} < n_{hBN}$, $n_{air} < n_{hBN}$), and the light coupling to these unwanted modes cannot be channeled into the collection optics, effectively lowering collection efficiency. Therefore, for optimal operation of the antenna, the hBN film thickness must be << $\lambda_{SPE} / (2 * n_{hBN})$. In practice, hBN layer thicknesses ≤ 20 nm are optimal to negate all unwanted channeling effects, making epitaxial hBN films, in this case ~ 10 nm thick, ideally suited for PDA integration. Figure 4b shows the simulated collection efficiency and relative power density from an in-plane linearly polarized dipole in hBN integrated with the PDA, calculated in accordance with previously published methods.[40] We find that more than 84% of the fluorescence emission can be channeled within a collection angle of 70°, which can be easily collected by using an oil immersion objective (NA of 1.46). This represents a minimum of a ~2-fold enhancement in the collection efficiency compared to measurements performed on Ni as in this work.

To quantify performance of the hBN/PDA device, we characterized the saturated count rate and purity of 3 representative SPEs. Figure 4c displays the saturation curves of 3 such SPEs. To provide an upper bound for the saturated emission rate from each SPE, we measured the count rate on the SPE and subtracted the background emission recorded from ~1µm away. Each recorded saturation curve was fit with the equation $f(P) = I_\infty \frac{P}{P_{Sat} + P}$, where $I_\infty$ is the saturated count rate, and $P_{Sat}$ is the saturation power. The obtained fits for SPE 1, 2, and 3 yield saturated count rates $I_\infty = 3.92\pm0.23*10^6$ cps, $2.27\pm0.05*10^6$ cps, and $2.59\pm0.19*10^6$ cps. The corresponding saturation powers are $P_{Sat} = 39.8\pm4.8$µW, $25.2\pm1.4$µW, and $36.0\pm5.6$µW respectively. The high-count rates, displaying average values $I_\infty$~$2.9*10^6$ cps at a very low saturation power of ~34 µW, demonstrate the promise of the integrated hBN/PDA device structure.

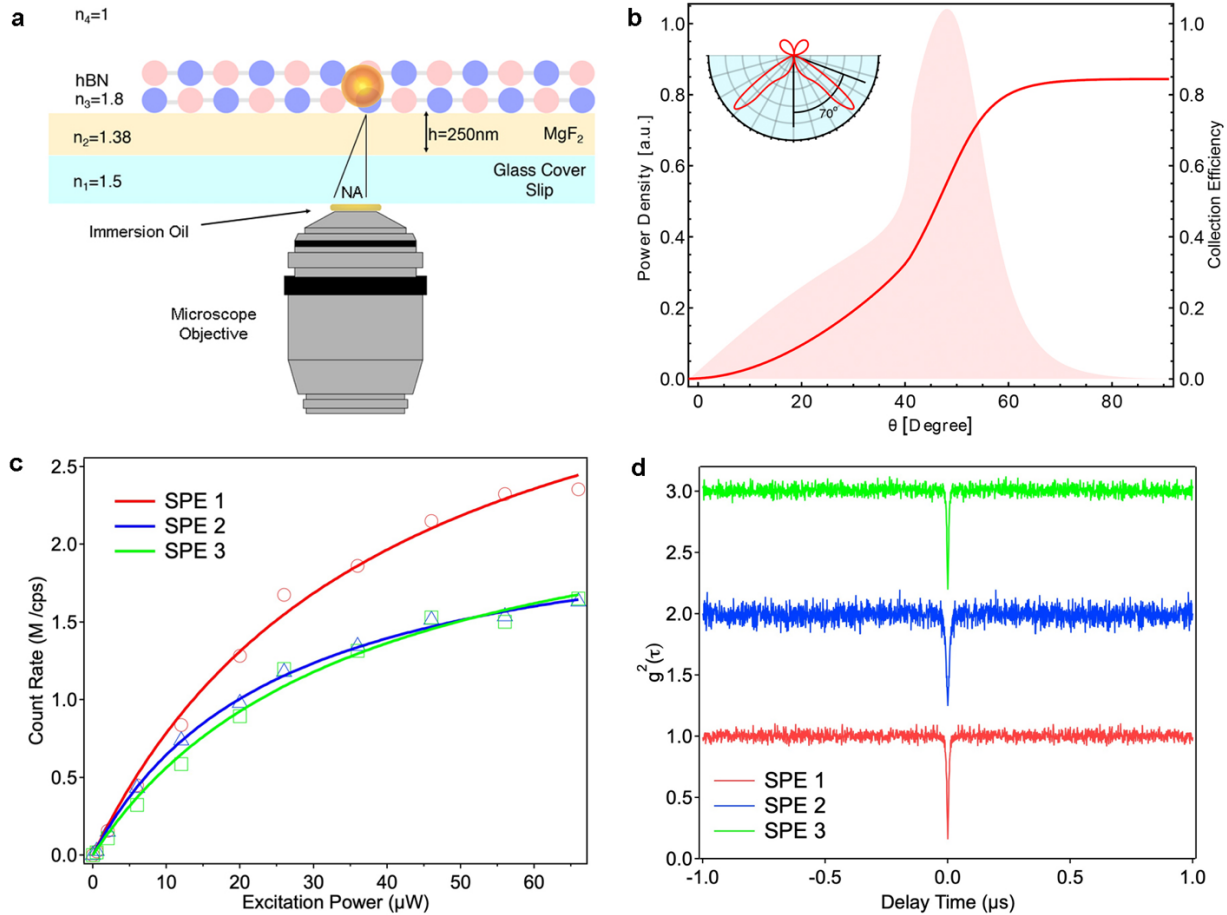

*Figure 4. Integration of hBN films with a planar dielectric antenna. **a.** Schematic representation of the hBN/PDA structure. **b.** Simulated power density (light red region) and collection efficiency (red line) as a function of the collection angle for a hBN SPE excited on the antenna. The maximum collection efficiency is ~84% at collection angles of ~70˚. The inset displays the simulated emission profile of a hBN SPE with a dipole oriented in plane **c.** Saturation curves for 3 separate hBN SPEs excited with a 532nm laser with powers from 0-66µW. **d.** Corresponding $g^2(\tau)$ collections for the 3 emitters in figure 3c.*

Figure 4d displays the associated $g^2(\tau)$ for SPE 1, 2, and 3, each collected with an excitation power of 10µW. We note in all cases, no additional spectral filtering or background subtraction was performed to provide a quantitative measurement of the emission purity. Fitting the collections for SPE 1, 2, and 3 yield $g^2(0)$=0.08, 0.18, and 0.07 respectively, confirming the high purity of the emission, see Supporting Information. The high-count rates, stability, and emission purity observed from the hybrid hBN/PDA device, provide a simple, yet promising and scalable platform for emerging quantum technologies.

In summary, we have demonstrated that hBN SPEs grown on polycrystalline Ni foil are incorporated preferentially in hBN on Ni (001) grains. We propose that the origin of the effect is variations in the diffusion and subsequent supply of atomic species during growth, a parameter known to vary with the grain orientation of the Ni catalyst. Intriguingly, controlling these diffusion effects on a nanometer scale could enable spatially controlled SPE incorporation during growth. These emitters display an enhanced fluorescence intensity (~2.5x) compared to those incorporated during hBN growth on Cu, with brightest showing ~7.6x10$^6$ counts/s. Finally, leveraging the nanoscale dimensions of the hBN thin films (~10 nm), we incorporate them in a planar dielectric antenna to maximize collection efficiency. Characterization of SPEs in the hybrid device yield average count rates of ~2.9*10$^6$ cps at a saturation power of ~ 34µW, and a high emission purity of ~ 90%. The hybrid devices are promising for room-temperature integrated quantum photonics with hBN and pose an efficient strategy to maximize collection efficiency from other two-dimensional quantum light sources.

**Supporting Information:**
The supporting information includes detailed experimental methods, as well as additional optical and materials characterization.

**Acknowledgments:**
We thank Simon White for help with wide-field imaging, Mark Lockrey for assistance with EBSD measurements, and Alexander Gumann from the TDSU1 nanofabrication facility (Max Planck Institute for the Science of Light) for magnesium fluoride deposition. We thank Vahid Sandoghdar for continuous support, and Carlo Bradac for fruitful discussions. We appreciate support from the Australian Research Council (DE170100169, DP180100077, DP190101058) and the Asian Office of Aerospace Research and Development ( FA2386-20-1-4014).

**References:**

1. Atatüre, M.; Englund, D.; Vamivakas, N.; Lee, S.-Y.; Wrachtrup, J. *Nature Reviews Materials* **2018,** 3, (5), 38-51.
2. Elshaari, A. W.; Pernice, W.; Srinivasan, K.; Benson, O.; Zwiller, V. *Nature Photonics* **2020**.
3. Toth, M.; Aharonovich, I. *Annu Rev Phys Chem* **2019,** 70, 123-142.
4. Liu, X.; Hersam, M. C. *Nature Reviews Materials* **2019**.
5. Peyskens, F.; Chakraborty, C.; Muneeb, M.; Van Thourhout, D.; Englund, D. *Nat Commun* **2019,** 10, (1), 4435.
6. Tonndorf, P.; Del Pozo-Zamudio, O.; Gruhler, N.; Kern, J.; Schmidt, R.; Dmitriev, A. I.; Bakhtinov, A. P.; Tartakovskii, A. I.; Pernice, W.; Michaelis de Vasconcellos, S.; Bratschitsch, R. *Nano Lett* **2017,** 17, (9), 5446-5451.
7. Kim, S.; Duong, N. M. H.; Nguyen, M.; Lu, T. J.; Kianinia, M.; Mendelson, N.; Solntsev, A.; Bradac, C.; Englund, D. R.; Aharonovich, I. *Advanced Optical Materials* **2019,** 7, (23).
8. Tran, T. T.; Wang, D.; Xu, Z. Q.; Yang, A.; Toth, M.; Odom, T. W.; Aharonovich, I. *Nano Lett* **2017,** 17, (4), 2634-2639.
9. Vogl, T.; Sripathy, K.; Sharma, A.; Reddy, P.; Sullivan, J.; Machacek, J. R.; Zhang, L.; Karouta, F.; Buchler, B. C.; Doherty, M. W.; Lu, Y.; Lam, P. K. *Nat Commun* **2019,** 10, (1), 1202.
10. Hoese, M.; Reddy, P.; Dietrich, A.; Koch, M. K.; Fehler, K. G.; Doherty, M. W.; Kubanek, A. *https://arxiv.org/abs/2004.10826* **2020**.
11. Li, X.; Scully, R. A.; Shayan, K.; Luo, Y.; Strauf, S. *ACS Nano* **2019,** 13, (6), 6992-6997.
12. Mendelson, N.; Chugh, D.; Reimers, J. R.; Cheng, T. S.; Gottscholl, A.; Long, H.; Mellor, C. J.; Zettl, A.; Dyakonov, V.; Beton, P. H.; Novikov, S. V.; Jagadish, C.; Tan, H. H.; Ford, M. J.; Toth, M.; Bradac, C.; Aharonovich, I. *https://arxiv.org/abs/2003.00949* **2020**.

13. Chejanovsky, N.; Mukherjee, A.; Kim, Y.; Denisenko, A.; Finkler, A.; Taniguchi, T.; Watanabe, K.; Dasari, D. B. R.; Smet, J. H.; Wrachtrup, J. https://arxiv.org/abs/1906.05903 **2019**.
14. Gottscholl, A.; Kianinia, M.; Soltamov, V.; Orlinskii, S.; Mamin, G.; Bradac, C.; Kasper, C.; Krambrock, K.; Sperlich, A.; Toth, M.; Aharonovich, I.; Dyakonov, V. *Nature Materials* **2020**.
15. Mendelson, N.; Xu, Z. Q.; Tran, T. T.; Kianinia, M.; Scott, J.; Bradac, C.; Aharonovich, I.; Toth, M. *ACS Nano* **2019,** 13, (3), 3132-3140.
16. Stern, H. L.; Wang, R.; Fan, Y.; Mizuta, R.; Stewart, J. C.; Needham, L. M.; Roberts, T. D.; Wai, R.; Ginsberg, N. S.; Klenerman, D.; Hofmann, S.; Lee, S. F. *ACS Nano* **2019,** 13, (4), 4538-4547.
17. Comtet, J.; Glushkov, E.; Navikas, V.; Feng, J.; Babenko, V.; Hofmann, S.; Watanabe, K.; Taniguchi, T.; Radenovic, A. *Nano Lett* **2019,** 19, (4), 2516-2523.
18. Hernández-Mínguez, A.; Lähnemann, J.; Nakhaie, S.; Lopes, J. M. J.; Santos, P. V. *Physical Review Applied* **2018,** 10, (4).
19. Lin, W.-H.; Brar, V. W.; Jariwala, D.; Sherrott, M. C.; Tseng, W.-S.; Wu, C.-I.; Yeh, N.-C.; Atwater, H. A. *Chemistry of Materials* **2017,** 29, (11), 4700-4707.
20. Abidi, I. H.; Mendelson, N.; Tran, T. T.; Tyagi, A.; Zhuang, M.; Weng, L. T.; Özyilmaz, B.; Aharonovich, I.; Toth, M.; Luo, Z. *Advanced Optical Materials* **2019**.
21. Feigelson, B. N.; Bermudez, V. M.; Hite, J. K.; Robinson, Z. R.; Wheeler, V. D.; Sridhara, K.; Hernandez, S. C. *Nanoscale* **2015,** 7, (8), 3694-702.
22. Lee, Y.-H.; Liu, K.-K.; Lu, A.-Y.; Wu, C.-Y.; Lin, C.-T.; Zhang, W.; Su, C.-Y.; Hsu, C.-L.; Lin, T.-W.; Wei, K.-H.; Shi, Y.; Li, L.-J. *RSC Adv.* **2012,** 2, (1), 111-115.
23. Gorbachev, R. V.; Riaz, I.; Nair, R. R.; Jalil, R.; Britnell, L.; Belle, B. D.; Hill, E. W.; Novoselov, K. S.; Watanabe, K.; Taniguchi, T.; Geim, A. K.; Blake, P. *Small* **2011,** 7, (4), 465-8.
24. Uchida, Y.; Nakandakari, S.; Kawahara, K.; Yamasaki, S.; Mitsuhara, M.; Ago, H. *ACS Nano* **2018,** 12, (6), 6236-6244.
25. Wigger, D.; Schmidt, R.; Del Pozo-Zamudio, O.; Preuß, J. A.; Tonndorf, P.; Schneider, R.; Steeger, P.; Kern, J.; Khodaei, Y.; Sperling, J.; de Vasconcellos, S. M.; Bratschitsch, R.; Kuhn, T. *2D Materials* **2019,** 6, (3).
26. Cho, H.; Park, S.; Won, D. I.; Kang, S. O.; Pyo, S. S.; Kim, D. I.; Kim, S. M.; Kim, H. C.; Kim, M. J. *Sci Rep* **2015,** 5, 11985.
27. Chou, H.; Majumder, S.; Roy, A.; Catalano, M.; Zhuang, P.; Quevedo-Lopez, M.; Colombo, L.; Banerjee, S. K. *ACS Appl Mater Interfaces* **2018,** 10, (51), 44862-44870.
28. Azzerri, N.; Colombo, R. L. *Metallography* **1976,** 9, (3), 233-244.
29. Blakely, J. M.; Mykura, H. *Acta Metallurgica* **1961,** 9, (1), 23-31.
30. Weston, L.; Wickramaratne, D.; Mackoit, M.; Alkauskas, A.; Van de Walle, C. G. *Physical Review B* **2018,** 97, (21).
31. Hite, J. K.; Robinson, Z. R.; Eddy, C. R., Jr.; Feigelson, B. N. *ACS Appl Mater Interfaces* **2015,** 7, (28), 15200-5.
32. Breitweiser, S. A.; Exarhos, A. L.; Patel, R. N.; Saouaf, J.; Porat, B.; Hopper, D. A.; Bassett, L. C. *ACS Photonics* **2019,** 7, (1), 288-295.
33. Nikolay, N.; Mendelson, N.; Özelci, E.; Sontheimer, B.; Böhm, F.; Kewes, G.; Toth, M.; Aharonovich, I.; Benson, O. *Optica* **2019,** 6, (8).
34. Lee, K. G.; Chen, X. W.; Eghlidi, H.; Kukura, P.; Lettow, R.; Renn, A.; Sandoghdar, V.; Götzinger, S. *Nature Photonics* **2011,** 5, (3), 166-169.
35. Chu, X.-L.; Götzinger, S.; Sandoghdar, V. *Nature Photonics* **2016,** 11, (1), 58-62.

36. Chu, X. L.; Brenner, T. J. K.; Chen, X. W.; Ghosh, Y.; Hollingsworth, J. A.; Sandoghdar, V.; Götzinger, S. *Optica* **2014,** 1, (4).
37. Riedel, D.; Rohner, D.; Ganzhorn, M.; Kaldewey, T.; Appel, P.; Neu, E.; Warburton, R. J.; Maletinsky, P. *Physical Review Applied* **2014,** 2, (6).
38. Chen, X. W.; Gotzinger, S.; Sandoghdar, V. *Opt Lett* **2011,** 36, (18), 3545-7.
39. Kim, S.; Froch, J. E.; Christian, J.; Straw, M.; Bishop, J.; Totonjian, D.; Watanabe, K.; Taniguchi, T.; Toth, M.; Aharonovich, I. *Nat Commun* **2018,** 9, (1), 2623.
40. Chen, X.-W.; Choy, W. C. H.; He, S. *Journal of Display Technology* **2007,** 3, (2), 110-117.